\documentclass[]{siamart0216}

\ifpdf
  \DeclareGraphicsExtensions{.eps,.pdf,.png,.jpg}
\else
  \DeclareGraphicsExtensions{.eps}
\fi

\newcommand{\TheTitle}{Extreme-scale Multigrid Components within PETSc} 
\newcommand{\TheAuthors}{D.~A. May, P. Sanan, K. Rupp, M.~G. Knepley and B.~F. Smith}

\headers{\TheTitle}{\TheAuthors}

\title{{\TheTitle}}

\author{
Dave A. May
       \thanks{Institute of Geophysics, ETH Z{\"u}rich, 
       Sonneggstrasse 5, 8092 Z{\"u}rich, Switzerland
       \email{dave.may@erdw.ethz.ch}}
\and
Patrick Sanan
       \thanks{Institute of Computational Sciences, 
       Universit{\`a} della Svizzera italiana, 
       Via Buffi 13, 6904 Lugano, Switzerland
       \email{patrick.sanan@usi.ch}}
\and
Karl Rupp
       \thanks{Linke Bahnzeile 7/6, 
       A-2486 Landegg, Austria
       \email{me@karlrupp.net}}
\and  
Matthew G. Knepley
       \thanks{Computational and Applied Mathematics, Rice University, 
       6100 Main MS-134
       Houston, TX 77005, USA
       \email{knepley@rice.edu}}
\and
Barry F. Smith
       \thanks{Mathematics and Computer Science Division, 
       Argonne National Laboratory, 
       9700 South Cass Avenue, 
       Argonne, IL 60439, USA
       \email{bsmith@mcs.anl.gov}}
}

\ifpdf
\hypersetup{
  pdftitle={\TheTitle},
  pdfauthor={\TheAuthors}
}
\fi

\usepackage{url}
\usepackage{algorithm,algpseudocode}
\usepackage{soul}
\usepackage{color}
\definecolor{cornsilk}{rgb}{1.0, 0.97, 0.86}
\usepackage{comment}
\usepackage{booktabs}
\usepackage{amsmath,amsfonts}
\usepackage{xcolor}

\algrenewcommand\algorithmicindent{1.0em} 
\newcommand{\pobj}[1]{\texttt{#1}}
\newcommand{\cmat}[1]{\boldsymbol{#1}}
\newcommand{\dmat}[1]{\mathbf{#1}}
\newcommand{\cvec}[1]{\boldsymbol{#1}}
\newcommand{\dvec}[1]{\mathbf{#1}}

\hypersetup{colorlinks=true,allcolors=red}
\hypersetup{citecolor=blue}
\hypersetup{urlcolor=cyan}
\usepackage{tcolorbox}

\begin{document}

\maketitle

\begin{abstract}
Elliptic partial differential equations (PDEs) frequently arise in continuum descriptions of physical processes relevant to science and engineering. 
Multilevel preconditioners represent a family of scalable techniques for solving discrete PDEs of this type and thus 
are the method of choice for high-resolution simulations.
The scalability and time-to-solution of massively parallel multilevel preconditioners can be adversely effected 
by using a coarse-level solver with sub-optimal algorithmic complexity.
To maintain scalability, agglomeration techniques applied to the coarse level have been shown to be necessary.

In this work, we present a new software component introduced within the 
Portable Extensible Toolkit for Scientific computation (PETSc) which permits agglomeration. 
We provide an overview of the design and implementation of this functionality, 
together with several use cases highlighting the benefits of agglomeration. 
Lastly, we demonstrate via numerical experiments employing geometric multigrid with structured meshes,
 the flexibility 
and performance gains possible using our MPI-rank agglomeration implementation.
\end{abstract}



\section{Introduction} \label{sec:intro}

In numerous branches of computational science and engineering, there is frequently a need to solve 
large systems of linear equations of the form
\begin{equation}
	\dmat A \dvec x = \dvec b, \quad \dmat A \in \mathbb R^{n \times n}, \quad \dvec x, \dvec b \in \mathbb R^n
	\label{eq:axb}
\end{equation}
which arise from the spatial discretization of partial differential equations (PDEs) which contain scalar (or vectorial) elliptic operators. 
Examples include, but are not limited to: steady-state thermal conduction
\begin{equation}
	- \nabla \cdot \bigl( k \nabla \phi \bigr) = f,
	\label{eq:poisson}
\end{equation}
where $k$ is the thermal conductivity and $\phi$ is the temperature; 
displacement $(\cvec u)$ formulations of small-strain elastostatics
\begin{equation}
	- \nabla \cdot \bigl( \bar{\cmat C} \, \epsilon[\cvec u] \bigr) = \cvec g,
	\label{eq:elasticity}
\end{equation}
where $\epsilon[\cvec x] = \tfrac{1}{2} \left( \nabla \cvec x + (\nabla \cvec x)^T \right)$ 
is the symmetric gradient of a vector and $\bar{\cmat{C}}$ is the constitutive tensor; 
and stationary incompressible Stokes flow
\begin{equation}
	- \nabla \cdot \bigl( 2 \eta \, {\epsilon}[\cvec v] \bigr) + \nabla p = \cvec h, \quad -\nabla \cdot \cvec v = 0,
	\label{eq:stokes}
\end{equation}
where $\cvec v, p$ is the velocity and pressure, respectively, and $\eta$ is the viscosity of the fluid.

Given the wide-spread availability of massively parallel, distributed memory computing capabilities 
offered by computing centres, application scientists continue to push the boundaries of 
both (i) simulation spatio-temporal resolution, and (ii) simulation throughput.
In the context of simulation resolution, leadership computing facilities provide resources which, if used at full capacity, 
can theoretically enable 3D simulations to be performed with billions or trillions of unknowns $(n)$ \cite{gmeiner2014parallel}. 
Alternatively, for a given spatio-temporal resolution, simulation throughput, or time-to-solution, 
can be accelerated by using more compute resources.
Whilst the algorithmic demands in the weak scaling limit (resolution) or strong scaling limit (throughput) are different, 
the most time-consuming part of application software which involve discretized elliptic operators invariably is associated 
with the solution of Eq.~\eqref{eq:axb}.

Preconditioned Krylov (iterative) methods are desirable solution algorithms in massively parallel computing environments as their 
fundamental building blocks, e.g.~matrix-vector products, norms, and dot products, readily map to distributed memory 
implementations. Nevertheless, without a suitable preconditioner, the number of iterations required by a Krylov method 
will rapidly increase under spatial refinement when applied to discrete elliptic operators.
To accelerate the convergence of Krylov methods, multilevel preconditioners such as those derived from two-level 
domain-decomposition methods (e.g.~additive Schwarz), algebraic multigrid, or geometric multigrid are preferred choices. 
These methods represent a family of efficient and scalable techniques for solving elliptic PDEs by eliminating errors 
across all  scales through a hierarchy of coarse meshes, or coarse subspaces. 
Selecting the particular multilevel preconditioner is dependent on both the characteristic of the physical problem 
(e.g.~the nature of the coefficient in the elliptic operator) and specific details related to the spatial discretization 
(e.g.~structured mesh versus unstructured, low order basis versus high order basis functions).
When geometric multigrid is a viable option, it will generally be more efficient than an algebraic 
multigrid implementation due to 
(i)  a scalable setup phase, 
(ii) the possibility to use highly optimized matrix-free smoothers on all levels, 
and (iii) the ability to utilize an identical stencil (non-zero structure) throughout all coarse-level operators. 
In this work, we are primarily concerned with multilevel preconditioners which utilize geometric 
information (such as a mesh) and thus we will focus the remainder of this discussion on such techniques. 

Geometric multigrid with re-discretized operators is the most common form of multilevel preconditioner used to solve 
elliptic problems when a hierarchy of successively refined meshes, and the interpolation and prolongation operations between them 
is readily available. 
The reason why multigrid is undoubtedly \textit{the preconditioner of choice} for elliptic problems 
is because the method is both algorithmically scalable (e.g. converges in a fixed number of 
iterations independent of the mesh resolution) and optimal. 
That is, for a given solution accuracy, the time-to-solution 
is proportional to the number of unknowns $n$, as are the storage requirements of the method. 
This is in contrast to, for example, sparse direct methods which have a time-to-solution which 
scales like $O(n^{3/2})$ (2D) or $O(n^2)$ (3D)  \cite{li2007use}.
The sequential multigrid preconditioner obtains its $O(n)$ behaviour as the amount of work 
to be performed on each level $k = 1, \dots, N$ (except the coarsest) is proportional to $n_k$. 
In the case of a three-dimensional problem with a coarsening factor of 2, the work per level decreases by a factor of $8$. 
On the coarsest level $(k=1)$ it is traditionally advocated to utilize an exact LU factorization. 
Despite the $O(n_1^2)$ scaling for the factorization, $n_1$ on the coarse level  
will contain $1/8^{N-1}$ times fewer unknowns than the finest level and thus the cost of the direct solve is negligible.   

\begin{figure}[h!]
\centering
\includegraphics[width=0.8\textwidth]{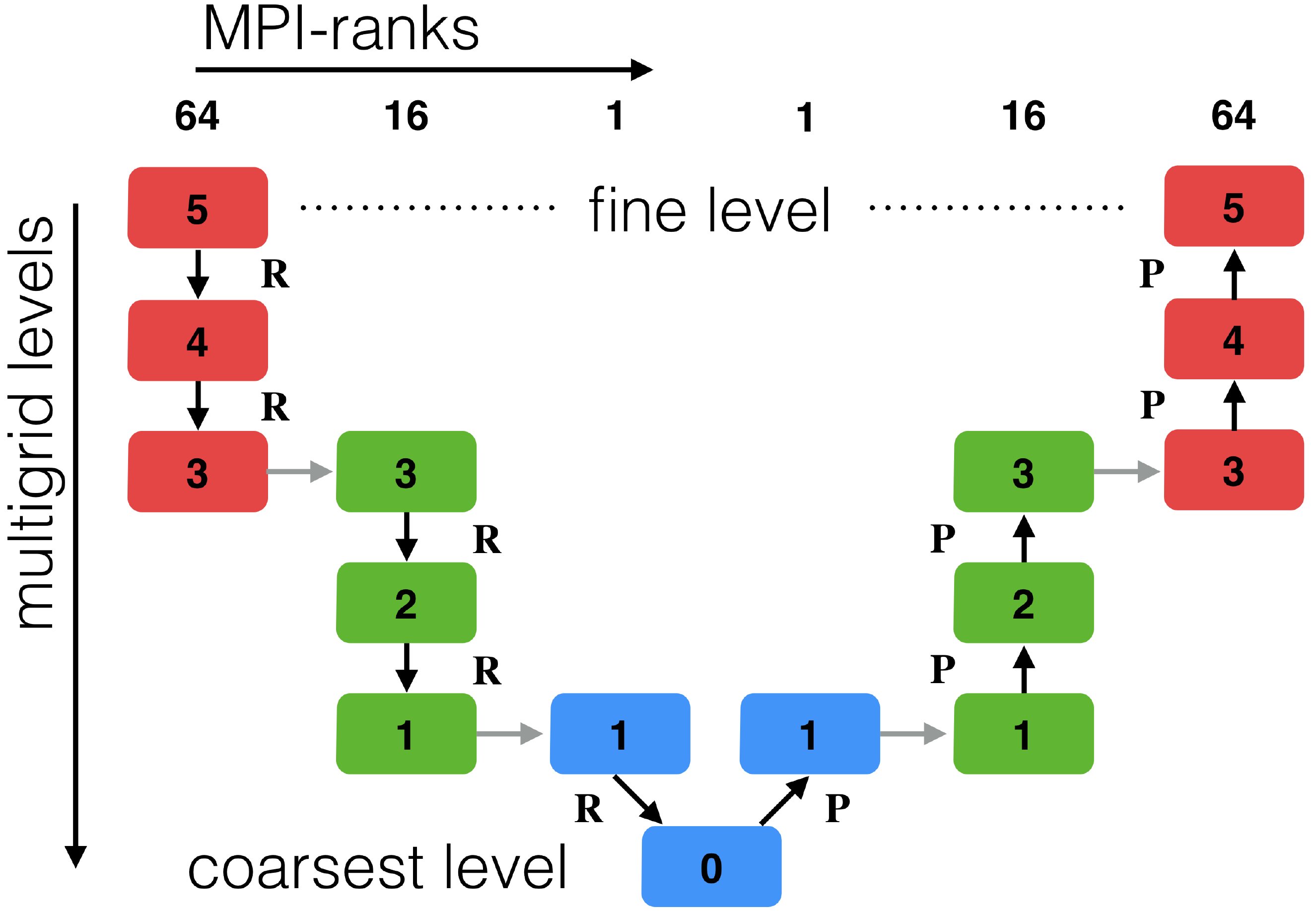}
\caption{Multigrid V-cycle with agglomeration. $\dmat R$ and $\dmat P$ denote the restriction and prolongation operators respectively. 
Grey horizontal arrows indicate where agglomeration occurs and data is moved between different communicators. \label{fig:agglom}}
\end{figure}

The treatment of coarse levels within the geometric multigrid hierarchy in a massively parallel distributed memory setting
requires some consideration as the ratio 
between communication and computation becomes larger with each coarser level.
Thus, the time spent on each level is not guaranteed to be a factor of $1/8$ times less 
than the next finest level. 
Moreover, the cost of performing an exact LU factorization of a 
very small problem distributed over many MPI-ranks may no longer be negligible. 
Several strategies have been proposed to treat coarse-level solves in multigrid preconditioners \cite{trottenberg2000multigrid}:
\begin{enumerate}
\setlength\itemsep{0em}
\item Truncate the number of multigrid levels when the cost of the communication cannot be 
overlapped with the computation on a given level, or when there is less than one unknown per MPI-rank. 
The depth of the multigrid hierarchy might be shallower than the equivalent sequential hierarchy, and thus 
an iterative solver on the ``coarsest'' level should be used. 
In practice it is observed that an inexact coarse-level solve up to a specified tolerance can yield an optimal 
multigrid preconditioner \cite{may2015scalable,trottenberg2000multigrid}. 
However, the cost of obtaining a fixed-precision inexact solve could be $n_k \log n_k$ 
and given the shallow nature of the hierarchy, this cost may be sufficient to degrade the expected $O(n)$ time-to-solution.
\item Allow for a subset of MPI-ranks to have zero unknowns, thereby introducing idle processing units. 
\item Agglomerate unknowns onto a new MPI communicator with fewer ranks. 
That is, coarsen the size of the MPI communicator in conjunction with the mesh coarsening to provide 
a more favourable balance between communication and computation. 
This concept is illustrated in Fig.~\ref{fig:agglom}.
\end{enumerate}

In practice, selecting the ideal coarse-level solver strategy is very much 
problem-dependent (e.g. constant coefficient versus spatially heterogenous) 
and machine-dependent (e.g. the latency associated of the network and the 
cost of global reductions versus floating point speed).

There are numerous examples in the literature where the three different coarse-level solver strategies have been adopted.
The truncation strategy is most frequently adopted. It is the simplest to implement 
as only the finest level is required to be partitioned over a single MPI communicator. 
In \cite{gmeiner2014parallel, gmeiner2015performance, gmeiner2015towards} the effect of using 
the conjugate gradient method as a coarse-level solver 
and its influence on the parallel scalability of multigrid was examined. On large-scale computations employing more than 100k processing cores, 
the cost of the coarse-level solve represented $\sim$15\% of the total compute time. 
Hierarchical Krylov methods \cite{mcinnes2014hierarchical} were utilized as the coarse level solver in \cite{may2015scalable} to reduce the number of global reductions required, and thereby providing a better balance between communication and computation. 


In \cite{Adams:2004:UIF:1048933.1049978, sundar2012parallel}, a fixed size MPI communicator was used across all levels and 
the number of active processors was reduced by completely eliminating unknowns from some MPI-ranks.  
A detailed examination of the performance gains of using this approach was not discussed by either author.
The downside of introducing ``zero work ranks'' is twofold. 
Firstly, \textit{if} any collective calls are used, all ranks must participate in the operation. 
Secondly, this programming model lacks generality beyond its applicability for linear algebra objects such as matrices and vectors. 
Using different MPI communicators is the correct paradigm to adopt and thus it naturally works with all parallel objects. 
Thus, agglomeration employing separate communicators is preferable in a general framework.

Various implementations of process agglomeration have been utilized within the solver 
community, e.g. \cite{blatt2012massively, emans2011coarse, may2015scalable, muller2014massively, reiter2013massively, rudi2015extreme}. 
In general, the agglomeration methods require a-priori specification of when agglomeration should occur 
and how aggressive it should be. However, based on most published results, there appears to be a lack of a performance 
model to guide such choices. Instead, experimentation is used to determine the minimum time-to-solution.
Consequently, the true benefits of using agglomeration are not clearly characterized or examined in detail in most studies. 
We note that this may in part be due to the problem- and machine-dependent nature of the scalability issues 
connected with coarse-level solvers.
For a range of problems containing less than about $3$M unknowns and scaling using up to 32 cores, a comparison of using a parallel LU factorization, an iterative method, and an agglomeration strategy together with algebraic multigrid has indicated that 
agglomeration is a superior approach \cite{emans2011coarse}. 
For medium-sized problems containing approximately $25$M unknowns on 4096 cores, the benefits of using agglomeration compared to an iterative 
coarse-level solver combined with smoothed aggregation algebraic multigrid on 4096 cores was demonstrated in \cite{may2015scalable}, 
with an overall solver speed-up of 1.8 being reported. 
For large-scale computations using the UG framework \cite{reiter2013massively} with problem sizes in excessive of 1B unknowns,  agglomeration was found essential to obtain good scalability past 4096 cores. In the example presented, solve 
times were a factor of 1.8 faster than those without agglomeration.

\subsection{Contributions}
In this work, we describe a newly developed component within the Portable Extensible Toolkit for 
Scientific computation (PETSc) \cite{petsc-user-ref,petsc-web-page,petsc-efficient} which permits agglomeration. 
Rather than develop a new multigrid implementation which supports agglomeration, we 
developed a flexible and re-usable component which can be utilized in a multitude of different ways.
The agglomeration operation is exposed within a new preconditioner object and as such can be readily composed with 
all other existing non-linear solvers, linear solvers, and preconditioner objects within PETSc. 
Unique to the agglomeration implementation we describe here is that it utilizes geometric information which 
may have been attached to the outer Krylov method. Therefore, it can be seamlessly used together with domain-decomposition 
and multigrid-type preconditioners.
The details pertaining to the design and implementation of the agglomeration preconditioner within PETSc is discussed, 
together with a number of typical use cases where this methodology can be beneficial. 
Lastly, we present several numerical experiments employing geometric multigrid to highlight both the flexibility of the proposed 
preconditioner, and the performance gains possible through MPI-rank agglomeration.

\section{Solvers and Discretization Components in PETSc}

\subsection{Preconditioned Krylov Subspace Methods}

Stationary, fixed-point iterative methods and preconditioned Krylov subspace methods within PETSc 
are defined by the \pobj{KSP} abstract class. The \pobj{KSP} object provides a rich family of iterative methods such as
Richardson, Chebyshev (fixed point); CG, GMRES (Krylov); GCR, FGMRES (flexible Krylov methods); and pipelined variants of CG and GMRES.

Essential to the convergence of Krylov methods is the choice of a preconditioner.
PETSc provides a large number of preconditioners via the \pobj{PC} class. This includes classical methods such as
 Jacobi, incomplete LU factorization (ILU), successive over-relaxation (SOR), block-Jacobi;
domain decomposition methods such as additive Schwarz (ASM), 
balancing Neumann-Neumann (NN), 
balancing domain decomposition by constraints (BDDC);
multilevel methods such as smoothed aggregation algebraic multigrid (GAMG); 
and a method for ``block'' or ``physics-based'' preconditioners (FieldSplit).


A fundamental design choice within PETSc is that solvers and preconditioners can be configured
at run-time. The degree of configurability ranges from generic solver parameters (e.g.~tolerances 
for stopping conditions), to the specific Krylov method and the type of preconditioner used. 
Moreover, solvers and preconditioner objects readily can be composed with each other at run-time 
using command line arguments (or an input file consisting of a set of command line arguments). 
Configuration and composability of nested solvers and preconditioners is enabled through the 
implementations assigning names (prefixes) to any internal \pobj{KSP} or \pobj{PC} objects.
These prefixes are concatenated together to provide unique textual identifiers for each configurable 
parameter which can be defined at run-time.
This mechanism allows end users to switch, at run-time, between a simple, non-scalable solver to a 
highly sophisticated, scalable method without changing a single line of code in their application software
\footnote{Consequently the PETSc acronym could be regarded as also being the Portable, Extensible Toolkit for \textit{Solver Composability}}.
For example, given an assembled matrix, users can either select to use CG with block-Jacobi and employ 
ILU on each sub-domain, or they can select to use CG with algebraic multigrid.

The degree of composability supports the end users' diverse and ever-changing requirements. 
A priori, the end user is unlikely to know the optimal solver and preconditioner configuration they will require. 
The choice of the ``optimal preconditioner'' is dependent on many factors. For instance, assuming the matrix 
is associated with a discretization of a PDE, influencing factors may include (but not exclusively): 
the characteristics of the underlying PDE (e.g.~elliptic versus parabolic), 
the nature of the coefficients in the PDE (e.g.~constant, smooth, discontinuous), 
the type of boundary conditions (e.g.~Dirichlet versus Neumann), 
the type of discretization, etc.
Changing any of these factors within the application software will invariably 
mandate changing the solver and preconditioner configuration in order to preserve the ``optimal'' choice.

\subsubsection{Multigrid}
To introduce the base multigrid implementation in PETSc (\pobj{PCMG}), we begin by 
first summarizing the classic two-level multigrid algorithm in Alg.~\ref{alg:mg2level} 
as applied to Eq.~\eqref{eq:axb}.
\begin{algorithm}
\caption{Two-level Multigrid}\label{alg:mg2level}
\begin{algorithmic}[1]
\State{Given $\dmat A$, $\dmat M$, $\dmat R$, $\dmat P$, $\dmat A_c$, $\dvec b$}
\State{Choose ${\dvec u}^0$}
\Repeat
\State ${\dvec u}^i \leftarrow {\dvec u}^i + {\dmat M}^{-1}\,({\dvec b} - \dmat A \dvec u^i)$
\Comment{pre-smooth $m$ times}
\State ${\dvec r} = {\dmat R}\,({\dvec b} - \dmat A \dvec u^i)$
\Comment{restrict residual}
\State ${\dmat A_c}\,{\dvec e} = {\dvec r}$
\Comment{solve for coarse grid correction} 
\State ${\dvec u}^i \leftarrow {\dvec u}^i + {\dmat P}\, {\dvec e}$
\Comment{prolongate error} 
\State ${\dvec u}^i \leftarrow {\dvec u}^i +{\dmat M}^{-1} ({\dvec b} - {\dmat A}\,{\dvec u}^i )$
\Comment{post-smooth $m$ times}
\State ${\dvec u}^{i+1} \leftarrow {\dvec u}^i$
\Comment{update for next iteration}
\Until{converged}
\end{algorithmic}
\end{algorithm}

A multigrid algorithm defines a hierarchy of levels -- with the top and bottom being referred to as ``fine'' 
and ``coarse'', respectively.
On each level (except the coarsest), we require 
(i) an operator $\dmat A$, and an operator used for preconditioning $\dmat M$,
(ii) a ``smoother'',
(iii) an operator to restrict a solution vector to the next coarsest level ($\dmat R$), and
(iv) an operator to prolongate a solution vector from the coarse level below ($\dmat P$).
On the coarsest level in the hierarchy we require a ``coarse'' level solver.
In the context of classical geometric multigrid, the ``smoother'' typically defines a fixed point iterative method which, 
when applied to a vector, removes high frequency error components, whilst the coarse-level solver is 
generally taken to be an LU factorization (exact solve).
In \pobj{PCMG}, both the smoother and the coarse-level solver are defined as \pobj{KSP} objects. 
As such, they can be configured to define a fixed-point method such as Richardson - Jacobi (classical smoother), or 
an exact solver. To enable different run-time configuration of the \pobj{KSP} on coarsest level and all other levels, the 
option prefix \pobj{-mg\_coarse} and \pobj{-mg\_levels} is used.


\subsection{Distribution Manager}
Solvers based purely on provided matrix entries are limited 
in their ability to perform well since one cannot take 
advantage of geometric or modeling information. Hence, a 
solver framework that allows access to this information is 
vital. The difficulty is creating a flexible, hierarchical way 
to provide this information that is nonintrusive, yet powerful. 
One unique feature of PETSc is the Distribution Manager (\pobj{DM}) 
abstract class, which provides information to the algebraic 
solvers regarding the 
mesh and physics but does not impose constraints on their 
management. We emphasize that \pobj{DM} is not a mesh management 
class and does not provide an interface to low-level mesh 
functionality; rather, \pobj{DM} provides an interface for accessing information 
relevant to and needed by the solver.
To that end, the \pobj{DM} class in PETSc provides a high-level
interface for obtaining mesh information to the solver.

A \pobj{DM} encodes two linear spaces, the \textit{global} space which encompasses
the entire problem (e.g.~as required to store the solution of a PDE) which is
appropriate for global solves, and the \textit{local} space composed of overlapping,
or ghosted, subspaces appropriate for local function evaluations. The \pobj{DM} can
create a vector from either space, and also a properly preallocated matrix from the
global space. In addition, it can provide a mapping between the global representation
of a field and its local representation, or vice versa.

The \pobj{DM} also establishes a natural notion of hierarchy. The user can obtain a
\textit{refined} or \textit{coarsened} version of the \pobj{DM}, along with operators
mapping fields between these spaces. While for structured grids (\pobj{DMDA}) these
operations can be defined solely in PETSc, for unstructured grids (\pobj{DMPlex}) we
use third party mesh manipulation packages such as Pragmatic~\cite{RokosGorman13}, Triangle~\cite{shewchuk96, triangle:homepage},
and TetGen~\cite{tetgen:homepage,Si2015} to produce refined and coarsened
meshes.

In addition to the hierarchy, the \pobj{DM} has an interface for creating subspaces. It provides
a consistent naming and representation of sub-fields of the global field using index sets, or
\pobj{IS} objects. By representing subspaces as simple sets of integers, interaction with
the solvers and linear algebra methods is simple and clean. A restriction operation is provided
onto the subspace, whether it be a subset of the fields or a subset of the domain. In fact,
a sub-\pobj{DM} can be created for the subspace, so the full problems can be solved consistently.



A \pobj{DM} can also be attached to a Krylov method and its preconditioner. 
Several preconditioner implementations in PETSc are ``\pobj{DM} aware'' and can use these objects in 
implementation specific manners. For example, the additive Schwarz preconditioner can use an attached \pobj{DM} 
to define overlapping sub-domains. Similarly, \pobj{PCMG} can use an attached \pobj{DM} to define both a mesh 
hierarchy through coarsening, and the restriction and prolongation operators between each mesh level. 

\section{Design \& Implementation of Telescope}

\subsection{Design Considerations}
The \pobj{DM} class currently does not provide support for repartitioning.
Even if it did, integrating repartitioning within the multigrid framework would be 
awkward, and the resulting functionality could not be directly used with other solver components. 
We decided that a more natural and re-usable way to introduce repartitioning into the  
composable solver space provided by PETSc is to embed this 
functionality \textit{within} a new preconditioner implementation -- we
call this implementation \pobj{Telescope}.

The general definition of a preconditioner in PETSc is the operation $\dvec y = \dmat A^{-1} \dvec x$.
From this, the essential philosophy behind \pobj{Telescope} is summarized below: 
\begin{enumerate}
\setlength\itemsep{0em}
\item Given an MPI communicator $\mathcal C$, create a new communicator $\mathcal C'$.
\item Repartition the input matrix $\dmat A$ and vector $\dvec x$ onto $\mathcal C'$, yielding $\dmat A'$ and $\dvec x'$. 
\item Apply a Krylov method to solve $\dmat A' \dvec y' = \dvec x'$ on $\mathcal C'$.
\item Scatter the solution $\dvec y'$ to $\mathcal C$ to obtain $\dvec y$.
\end{enumerate}
The notion of using a ``preconditioner'' as an entry point to enable repartitioning or modification of a matrix 
is utilized in the existing PETSc preconditioners \pobj{Redundant} and \pobj{Redistribute} already.


The size (number of MPI-ranks) of the communicator $\mathcal C'$, denoted by $n_{\mathcal C'}$, 
is defined by $n_{\mathcal C} / r$, where $r$ is the rank reduction factor specified by the user. 
A subset of MPI-ranks from $\mathcal C$ are used to define $\mathcal C'$.
Denoting the index of each MPI-rank in $\mathcal C$ via $R_{\mathcal C}^k \in [0,n_{\mathcal C}-1]$, 
any index for which  $\text{mod}(R_{\mathcal C}^k , r)$ equals zero is included within $\mathcal C'$.
The strided layout of MPI-ranks in $\mathcal C'$ is advantageous when the available RAM per-core 
is limited and thus distributing the repartitioned objects $\dmat A', \dvec x'$ over more compute nodes is desirable.
MPI-ranks in $\mathcal C$ which are not used within $\mathcal C'$ are \textit{idle} during the 
application of the nested Krylov method. 
We argue that this choice is both more efficient and simpler to implement than an alternative implementation 
which performs redundant calculations. 

We note that objects defined on both communicators $\mathcal C$ and $\mathcal C'$ (e.g.~$\dmat A, \dvec x, \dvec y$) are held in memory.

\subsection{Special Cases}
Here we describe the different features of the implementation of \pobj{Telescope} which will allude
to its potential usage within the context of multilevel and domain decomposition preconditioners, and highlight
the differences with the related preconditioner \pobj{Redundant}.
\begin{itemize}
\setlength\itemsep{0em}
	\item[C1] In PETSc, a sparse matrix with communicator $\mathcal C$ will be partitioned across 
	all MPI-ranks using a row-wise decomposition, e.g.~
	$\dmat A = [ \dmat A_0; \, \dmat A_1; \, \dots; \, \dmat A_{n_{\mathcal C}-1}]$.
	When \pobj{Telescope} is provided with $\dmat A$ and a \pobj{DM} has not been 
	attached to the preconditioner, the repartitioned matrix $\dmat A'$ is constructed by 
	contiguously fusing rows of $\dmat A$ such that
	$\dmat A' = [ \dmat A'_0; \, \dots; \, \dmat A'_{n_{\mathcal C'}-1}]$ where
	$$
		\dmat A'_k = \big[ \dmat A_{kr}; \,\dmat  A_{kr+1}; \, \dots; \, \dmat A_{kr + r-1} \big],
	$$ 
	$k$ is the index of an MPI-rank within $\mathcal C'$ and $r$ is the user-specified rank reduction factor.
	The redistribution of $\dmat A$ occurs in two phases which are depicted in Fig.~\ref{fig:matredist}b), c).	
	The intermediate matrices $\dmat A^r_k$ are sequential objects that are formed through a combination of copying 
	local data from the same rank (blue arrows) and scattering data from ranks being agglomerated (red arrows).
	The operator $\dmat A'$ is identical to that which would be defined using \pobj{Redundant}, 
	with the exception, that it is only defined on a subset of MPI-ranks within the parent communicator ${\mathcal C}$
	and no redundant work is performed on the other MPI-ranks.
\begin{figure}[h!]
\centering
\includegraphics[width=0.8\textwidth]{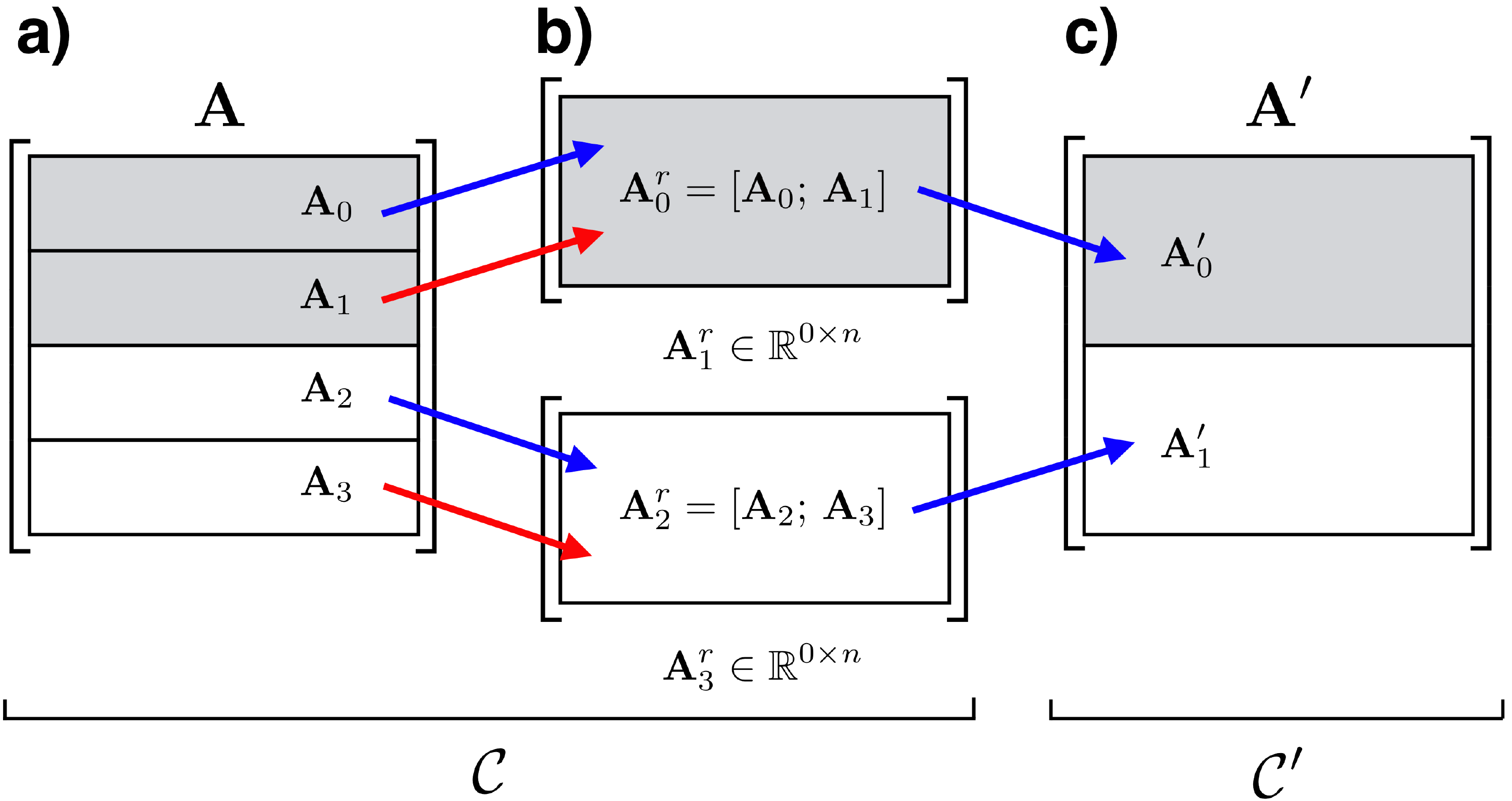}
\caption{Matrix redistribution: a) Original matrix defined on $\mathcal C$; b) Intermediate sequential matrices on all ranks in $\mathcal C$. Ranks not in $\mathcal C'$ contain zero rows; c) Redistributed matrix defined on the reduced communicator $\mathcal C'$. \label{fig:matredist}}
\end{figure}

	\item[C2] PETSc provides two mechanisms to remove null spaces from an operator.
	The user can provide either (or both): 
	  a set of $n_s$ vectors $\pmb \phi_i$ representing each null space, or
	  a function which will perform the removal.
	If the original operator $\dmat A$ has a null space attached, the vectors $\pmb \phi_i$ are scattered onto $\mathcal C'$
	and attached to $\dmat A'$. Any user-provided null space removal function attached to $\dmat A$ 
	is also assigned to the null space defined on $\dmat A'$. 
	\pobj{Redundant} currently does not support propagation of the null space to the repartitioned matrix.
	\item[C3] To support scalable multilevel algorithms, \pobj{Telescope} exploits geometric 
	and discretization information which is provided by an attached \pobj{DM} defined on $\mathcal C$. 
	We denote by $\mathcal T$ the discretization defined by the \pobj{DM}.
	An exact definition of $\mathcal T$ cannot be provided as it depends on the type of the \pobj{DM}, 
	but it should be regarded as representing: geometric primitives (e.g. points, edges, faces, cells);
	topological relationships between the geometric primitives; 
	geometry (e.g. coordinates); and field information (e.g. number of degrees of freedom attached to each geometric primitive).
	For example, in the case of a \pobj{DMDA} (structured grid), $\mathcal T$ defines the number of grid 
	points in each $i,j,k$ direction in both the local and global space, and the coordinates of each grid point. 
	Due to the  structured nature of the \pobj{DMDA}, $\mathcal T$ also implicitly defines the ordering of unknowns and the 
	connectivity between each grid point.
	
	When a \pobj{DM} is provided, $\mathcal T$ is repartitioned onto $\mathcal C'$, resulting in $\mathcal T'$. 
	The redistribution phase of $\dmat A$ and $\dvec x$ must preserve the re-ordering of the unknowns which occurred 
	when $\mathcal T$ was repartitioned. 
	 	
	Currently, \pobj{Telescope} only supports the repartitioning of 2D and 3D \pobj{DMDA}s. However, it is designed 
	in a modular fashion such that support for other \pobj{DM} implementations can be introduced. 
	From $\mathcal T$ we create a \pobj{DMDA} with identical coordinates (if defined), field, and 
	discretization properties on $\mathcal C'$. 
	We exploit the structured IJK topology of the \pobj{DMDA} and the predefined $i,j,k$ order of 
	both the MPI-ranks and the global unknowns adopted by the \pobj{DMDA} implementation to construct 
	the mapping between the ordering of the unknowns in $\mathcal T'$ and $\mathcal T$. 
	This mapping is expressed as an explicitly assembled permutation matrix $\hat{\dmat P}$ defined on $\mathcal C$. 
	
	Prior to the scatter of $\dvec x$ and the solve on $\mathcal C'$, the input vector is permuted according to $\hat{\dmat P}^T \dvec x$. 
	Similarly, after the solution $\dvec y'$ is scattered to $\mathcal C$, the inverse permutation $\hat{\dmat P} \dvec y$ is applied.
	Two methods exist to permute the operator $\dmat A$ so that the unknowns are ordered in a manner consistent with the  
	unknown ordering defined by $\mathcal T'$. We either (i) form 
	$\dmat A_p = \hat{\dmat P}^T \dmat A \hat{\dmat P}$ explicitly during the setup phase and redistribute it according to C1, 
	or (ii) if the user provided a callback function to assemble the operator\footnote{This is achieved by calling \pobj{KSPSetComputeOperators()}}, the user function is
	propagated to the sub-\pobj{KSP} and this function will be responsible for assembling $\dmat A'$ using $\mathcal T'$.	
	

	The decision to use an assembled permutation matrix stemmed from a lack of support for parallel vector permutations.
	Despite this, our choice yields a simple strategy for permuting the operator (optionally) and vectors, and furthermore, it is highly 
	efficient as it utilizes the optimized kernels within PETSc for \pobj{MatPtAP()} and \pobj{MatMult()}.
\end{itemize}

\subsection{Example Use Cases} \label{sec:usecases}
Here we elaborate on a number of potential use cases where the invocation of \pobj{Telescope} 
can be advantageous in the context of domain decomposition and multilevel preconditioners.
For the purpose of this discussion, we will consider the numerical solution of Laplace's equation 
using a finite-difference discretization with a mesh which is spatially decomposed 
across a communicator $\mathcal C$. 
For this problem, the natural preconditioner to employ is geometric multigrid. 
For each use case, we provide the relevant PETSc options to enable the configuration of each solver.
These solver options can be used with a standard PETSc example (\pobj{ex45}), which solves 
Laplace's equation in 3D and uses the \pobj{DMDA} to define the finite-difference grid.\footnote{
This example is provided with the PETSc source distribution and can be found  in the directory \pobj{src/ksp/ksp/examples/tutorials}}
\subsubsection{Multigrid with Truncation} \label{sec:truncation} 
	Suppose a user defines their own mesh hierarchy with $N$ levels, each of which is decomposed over a single MPI communicator $\mathcal C$.
	The definition and assembling  of the restriction, prolongation and coarse grid operators is performed by the user
	and these are subsequently provided to \pobj{PCMG}.  
	Due to current implementation restrictions, the mesh can only be coarsened until the coarse-level problem 
	contains $n_{\mathcal C}$ unknowns. 
	To enable an exact sequential LU factorization on the coarse grid, the problem can be algebraically 
	repartitioned onto $\mathcal C'$ using $r = n_{\mathcal C}$ with the following options:
\begin{tcolorbox}[colframe=red,colback=cornsilk,boxrule=0.5pt,arc=4pt,
      left=-6pt,right=6pt,top=6pt,bottom=6pt,boxsep=0pt]
	\begin{verbatim}
    -pc_type mg
    -pc_mg_levels N
    -mg_coarse_pc_type telescope
    -mg_coarse_pc_telescope_reduction_factor nc
    -mg_coarse_telescope_pc_type lu
	\end{verbatim}
\end{tcolorbox}	

\subsubsection{Repartitioned Coarse Grids}
	Assuming that the discretization for the Laplace equation was described via a \pobj{DMDA}, 
	the user only needs to provide the fine-level operator and attach the \pobj{DM} -- 
	from this information, a complete geometric multigrid hierarchy employing Galerkin 
	coarse-level operators can be constructed. 
	
	Suppose we wish to coarsen the \pobj{DMDA} until the coarse grid consists of 
	$n_x \times n_y \times n_z$ points in directions $i, j, k$, and then use geometric 
	multigrid as the preconditioner for this coarse problem with a communicator containing fewer ranks.
	This can be achieved at runtime by (i) recursively defining each ``coarsest'' 
	level \pobj{KSP}/\pobj{PC} to use \pobj{Telescope}, and (ii) configuring the sub-\pobj{KSP} within \pobj{Telescope} to use \pobj{PCMG}.
	We note that this is not an automated procedure and users must manually 
	determine the number of multigrid levels which can be defined on each communicator
	without the \pobj{DMDA} being over-decomposed. 
	Furthermore, the user must manually choose the number of multigrid levels within each 
	\pobj{Telescope} object in order to reduce the grid to the target size of $n_x \times n_y \times n_z$ points. Nevertheless,
    all these manual choices can be made at 
	run-time, thereby enabling performance tuning to be easily conducted.
	
	Below we provide options to define a single phase of repartitioning. Two stages of multigrid are invoked 
	in which Galerkin coarse operators are used throughout. The first stage of multigrid on $\mathcal C$ 
	employs $N_1$ levels, the second hierarchy on $\mathcal C'$ uses $N_2$ levels. 
	Note that the total number of multigrid levels in the fused hierarchy is $N_1 + N_2 - 1$.
\begin{tcolorbox}[colframe=red,colback=cornsilk,boxrule=0.5pt,arc=4pt,
      left=-6pt,right=6pt,top=6pt,bottom=6pt,boxsep=0pt]
	\begin{verbatim}
    -pc_type mg
    -pc_mg_levels N1
    -pc_mg_galerkin
    -mg_coarse_pc_type telescope
    -mg_coarse_pc_telescope_reduction_factor r
    -mg_coarse_telescope_pc_type mg
    -mg_coarse_telescope_pc_mg_levels N2
    -mg_coarse_telescope_pc_mg_galerkin
	\end{verbatim}
\end{tcolorbox}
\subsubsection{Hybrid Coarse Operator Construction}
	The convergence of geometric multigrid when applied to elliptic operators possessing a 
	coefficient structure which is highly heterogenous can be challenging. 
	The primary concern is how to best represent the coefficient structure on the coarse grids. 
	When the coefficients are highly variable, hybrid strategies which employ different techniques 
	to define the coarse-level operators have proven to yield improved time-to-solution with 
	minimal storage overhead \cite{may2015scalable,sundar2012parallel}.

	In the example below, we seamlessly combine re-discretized operators on $N_1-1$ levels, followed by $N_2-1$ levels 
	employing Galerkin coarse operators, and in the last phase we utilise smoothed aggregation multigrid 
	to define the operators on the remaining coarse levels.
\begin{tcolorbox}[colframe=red,colback=cornsilk,boxrule=0.5pt,arc=4pt,
      left=-6pt,right=6pt,top=6pt,bottom=6pt,boxsep=0pt]
	\begin{verbatim}
    -pc_type mg
    -pc_mg_levels N1
    -mg_coarse_pc_type telescope
    -mg_coarse_pc_telescope_reduction_factor r
    -mg_coarse_telescope_pc_type mg
    -mg_coarse_telescope_pc_mg_levels N2
    -mg_coarse_telescope_pc_mg_galerkin
    -mg_coarse_telescope_mg_coarse_pc_type gamg
	\end{verbatim}
\end{tcolorbox}
\subsubsection{Sub-Domain Smoothers with Constant Size} \label{sec:sscontsize}
	There are instances when a simple smoother (e.g. Chebyshev - Jacobi) does not 
	efficiently remove high frequency components from the residual. 
	In this situation, sub-domain smoothers defined via a block-Jacobi preconditioner 
	coupled with the application of ILU(0), or Gauss-Seidel, may be effective. 
	Modern computer architectures employ compute nodes which possess many cores (denoted by $r_n$), and 
	the immediate trend is that $r_n$ will continue to increase in the coming future.
	From an efficiency perspective of all operations, in particular vector operations and matrix-vector products, 
	it is desirable to use all $r_n$ cores per compute node. 
	
	Smoothers defined using block-Jacobi have the undesirable characteristic that the smoothing 
	properties are strongly connected with the size of the sub-domain. Thus, as $r_n$ becomes larger,
	these smoothers may cease to be beneficial. To that end, within the definition of the smoother, 
	we can invoke \pobj{Telescope} and request to coarsen the parent communicator size by a factor $r_n$, 
	thereby conserving the size of the sub-domain (independent of the number of cores per node) and 
	thus preserving the smoothing characteristics associated with using ILU(0) on the sub-domain. 
	The advantages of using techniques to maintain the size of the sub-domain in the 
	context of a Gauss-Seidel smoother have been previously demonstrated \cite{hoefler2013mpi}.
	The options below provide such a smoother configuration.
\begin{tcolorbox}[colframe=red,colback=cornsilk,boxrule=0.5pt,arc=4pt,
      left=-6pt,right=6pt,top=6pt,bottom=6pt,boxsep=0pt]
	\begin{verbatim}
    -pc_type mg
    -pc_mg_levels N
    -mg_levels_pc_type telescope
    -mg_levels_pc_telescope_reduction_factor rn
    -mg_levels_telescope_pc_type bjacobi
	\end{verbatim}
\end{tcolorbox}
\subsubsection{Smoothers with Different Spatial Decomposition}  \label{sec:ssdiffsd}

	Incomplete factorizations such as ILU(0) or ICC(0) can be a highly effective smoother 
	for problems which possess strong anisotropy arising due to rheological layering \cite{lechmann2011comparing}, or 
	from the underlying spatial discretization (e.g.~high aspect ratio elements, see Chapter 7 \cite{trottenberg2000multigrid}). 
	Such an approach was advocated in \cite{brown2013achieving, isaac.sisc.2015}, where ICC(0) with a 
	column-oriented (e.g.~perpendicular to the anisotropy) ordering of the unknowns  provides an exact column solve
	and thus is a highly efficient smoother. 
	Similar observations were reported in \cite{lechmann2011comparing}, where multigrid with 
	ILU(0) on structured meshes was found to produce robust convergence provided that mesh decomposition 
	in the direction of the gradient of the viscosity layering was avoided -- thereby mandating a $(d-1)$ spatial decomposition of the mesh.
	
	In practice, restricting the dimensionality of the spatial decomposition can degrade the overall parallelism possible and 
	adversely impact performance due to large surface area to volume ratios. 
	To avoid this issue, one can consider partitioning the 3D, structured mesh problem over $m \times n \times p$ MPI-ranks, 
	but require that the application of the smoother is performed on $m \times n \times 1$ ranks. 
	This can be invoked by specifying the spatial decomposition to be used by the repartitioned \pobj{DMDA} defined on $\mathcal C'$ 
	using the following options:
\begin{tcolorbox}[colframe=red,colback=cornsilk,boxrule=0.5pt,arc=4pt,
      left=-6pt,right=6pt,top=6pt,bottom=6pt,boxsep=0pt]
	\begin{verbatim}
    -pc_type mg
    -pc_mg_levels N
    -mg_levels_pc_type telescope
    -mg_levels_pc_telescope_reduction_factor r
    -mg_levels_telescope_repart_da_processors_z 1
	\end{verbatim}
\end{tcolorbox}

\section{Numerical Experiments}

Numerical experiments were performed on either ``Piz Daint'' or ``Edison''.
Piz Daint, located at the Swiss National Supercomputing Centre (CSCS),  is a Cray XC30 system 
with a total of 5,272 compute nodes, each equipped with an 8-core, 
64-bit Intel Sandy Bridge processor (E5-2670) and an Nvidia Tesla K20X GPU.
Piz Daint employs the Cray Aries high-speed interconnect with Dragonfly topology. 

Edison is a Cray XC30 system with a total of 5,576 compute nodes located at NERSC. 
Each compute nodes possesses two sockets, each equipped with a 12-core, 
64-bit Intel Ivy Bridge processor. 
Edison employs the Cray Aries high-speed interconnect with Dragonfly topology. 

\subsection{Agglomeration Profiling}


To profile the time required for (a) the setup phase ($T_\text{setup}$) and 
for (b) the vector permutation and scattering occurring within each application 
of the \pobj{Telescope} preconditioner ($T_\text{apply}$), we consider two different discretizations defined on a \pobj{DMDA} which we identify 
as \emph{Disc.~A} and \emph{Disc.~B}. 
\emph{Disc.~A} consists of a low-order 3D finite-difference discretization of the scalar Laplace equation (\pobj{ex45}).
The number of nodal points in the finite-difference mesh is $N^3$ and
the maximum number of non-zero entires per row in the operator is 7.
\emph{Disc.~B} consists of a stabilised, low-order ($Q_1$) 3D mixed finite-element discretization of a variable 
viscosity incompressible Stokes problem (\pobj{ex42})\footnote{This example is provided with the PETSc source distribution and can be found  in the directory \pobj{src/ksp/ksp/examples/tutorials}}. 
The total number of finite-elements used to discretize the domain is denoted by $M^3$. 
The $Q_1$ finite-element stencil contains 27 points, each with four unknowns ($u,v,w,p$), 
hence the maximum number of non-zero entries per row is 108.

Due to the particular configuration of \pobj{Telescope}, in all experiments performed here the 
operator $\dmat A$ was not required to be explicitly permuted by the setup phase of \pobj{Telescope}.
The reported values for $T_\text{setup}$ reflect the time required to: construct the communicator $\mathcal C'$;
create the \pobj{DMDA} on $\mathcal C'$; scatter the mesh coordinates from $\mathcal C \rightarrow \mathcal C'$; and 
assemble the permutation matrix $\hat{\dmat P}$.
All numerical results reported in this sub-section were performed on Piz Daint.


In Table~\ref{tab:telescopeA} we report the time required for the setup phase ($T_\text{setup}$) and the time required to perform both 
vector scatters ($\dvec x \rightarrow \dvec x', \dvec y' \rightarrow \dvec y$) and permutations ($T_\text{apply}$), for a range of 
different mesh sizes, communicator sizes ($n_{\mathcal C}$) and rank reduction factors ($r$) using \emph{Disc. A}.
Times reported are the \textit{maximum} over all ranks within $\mathcal C$.
We considered scenarios with a fixed sized sub-domain of $2^3$ grid points. 
For the large-scale tests utilizing 13k cores, the setup cost of \pobj{Telescope} is less than 50 ms.
Furthermore, the parallel permutations of the input and output vectors, and the 
two scatters to required to map the permuted input / output vector from $\mathcal C$ to $\mathcal C'$ (and vice versa), 
collectively require less than 600~$\mu$s when executed on more than 13k cores.
The setup cost is observed to be approximately 7 times larger than the cost of scattering the input / output vectors 
when $n_{\mathcal C} = 64$, and approximately 100 times larger when $n_{\mathcal C} = 13824$.

\begin{table}[h!]
\centering
\caption{Setup time ($T_\mathrm{setup}$) and application time ($T_\mathrm{apply}$) for Disc. A using different problem sizes ($N^3$ grid points), different numbers of MPI-ranks ($n_{\mathcal C}$), and reduction factors $r$.
See text for model definition.}
\begin{tabular}{r r r r r}
\toprule
$n_{\mathcal C}$     &$N$    &$r$    &$T_\text{setup}$ (s) &$T_\text{apply}$ (s)\\
\toprule
64                       &$8$         &8            &1.64E$-$03 &8.11E$-$05   \\
64                       &$8$         &16          &1.77E$-$03 &1.00E$-$04   \\
64                       &$8$         &32          &1.88E$-$03 &1.51E$-$04   \\
64                       &$8$         &64          &2.05E$-$03 &2.80E$-$04   \\
\midrule
4096                    &$32$      &8         &3.02E$-$02 &5.63E$-$04   \\
4096                    &$32$      &16       &3.82E$-$02 &3.84E$-$04   \\
4096                    &$32$      &32       &3.19E$-$02 &3.74E$-$04   \\
4096                    &$32$      &64       &3.12E$-$02 &6.21E$-$04   \\
\midrule
13824                  &$48$         &8            &4.37E$-$02 &4.30E$-$04   \\
13824                  &$48$         &16          &4.55E$-$02 &3.53E$-$04   \\
13824                  &$48$         &32          &5.76E$-$02 &5.58E$-$04   \\
13824                  &$48$         &64          &5.50E$-$02 &5.62E$-$04   \\
%
\bottomrule
\end{tabular}
\label{tab:telescopeA}
\end{table}

In Table~\ref{tab:telescopeB} we examine the influence of the rank reduction factor $r$ for a mesh of fixed size 
(sub-domains of $2^3$ elements), with respect to the setup and application phase 
of \pobj{Telescope} when applied to \emph{Disc. B}.
Times reported are the \textit{maximum} over all ranks within $\mathcal C$.
At low cores counts ($n_{\mathcal C} = 64$), the setup time was observed to be less than $\approx 20$~ms. 
With an increased number of MPI-ranks ($n_{\mathcal C}$), the setup cost grows, but remains less than $130$~ms.
The setup cost is observed to be approximately 4 times larger than the cost of scattering the input / output vectors 
when $n_{\mathcal C} = 64$, and approximately 20 times larger when $n_{\mathcal C} = 13824$. 
At the scale of 13824 cores, the application time appears to scale approximately linearly with respect to the stencil size cf. 
with the finite-difference results in Table~\ref{tab:telescopeA}.

\begin{table}[h!]
\centering
\caption{Setup time ($T_\mathrm{setup}$) and application time ($T_\mathrm{apply}$) for Disc. B using different problem sizes ($M^3$ finite elements), different numbers of MPI-ranks ($n_{\mathcal C}$), and reduction factors $r$. 
See text for model definition.}
\begin{tabular}{r r r r r r}
\toprule
$n_{\mathcal C}$     &$M$    &$r$    &$T_\text{setup}$ (s) &$T_\text{apply}$ (s)\\
\midrule
64                       &$8$         &8            &6.34E$-$03 &1.39E$-$03   \\
64                       &$8$         &16          &1.02E$-$02 &2.06E$-$03  \\
64                       &$8$         &32          &1.23E$-$02  &3.26E$-$03  \\
64                       &$8$         &64          &1.72E$-$02 &4.44E$-$03    \\
\midrule
4096                   &$32$         &8            &3.96E$-$02 &1.53E$-$03   \\
4096                   &$32$         &16          &4.93E$-$02 &2.58E$-$03   \\
4096                   &$32$         &32          &5.76E$-$02 &4.20E$-$03   \\
4096                   &$32$         &64          &7.39E$-$02 &7.33E$-$03   \\
\midrule
13824                 &$48$         &8            &8.04E$-$02 &1.58E$-$03   \\
13824                 &$48$         &16          &8.91E$-$02 &2.60E$-$03   \\
13824                 &$48$         &32          &1.02E$-$01 &4.20E$-$03   \\
13824                 &$48$         &64          &1.30E$-$01 &7.37E$-$03   \\
\bottomrule
\end{tabular}
\label{tab:telescopeB}
\end{table}

\begin{figure}[h!]
\centering
\includegraphics[width=0.8\textwidth]{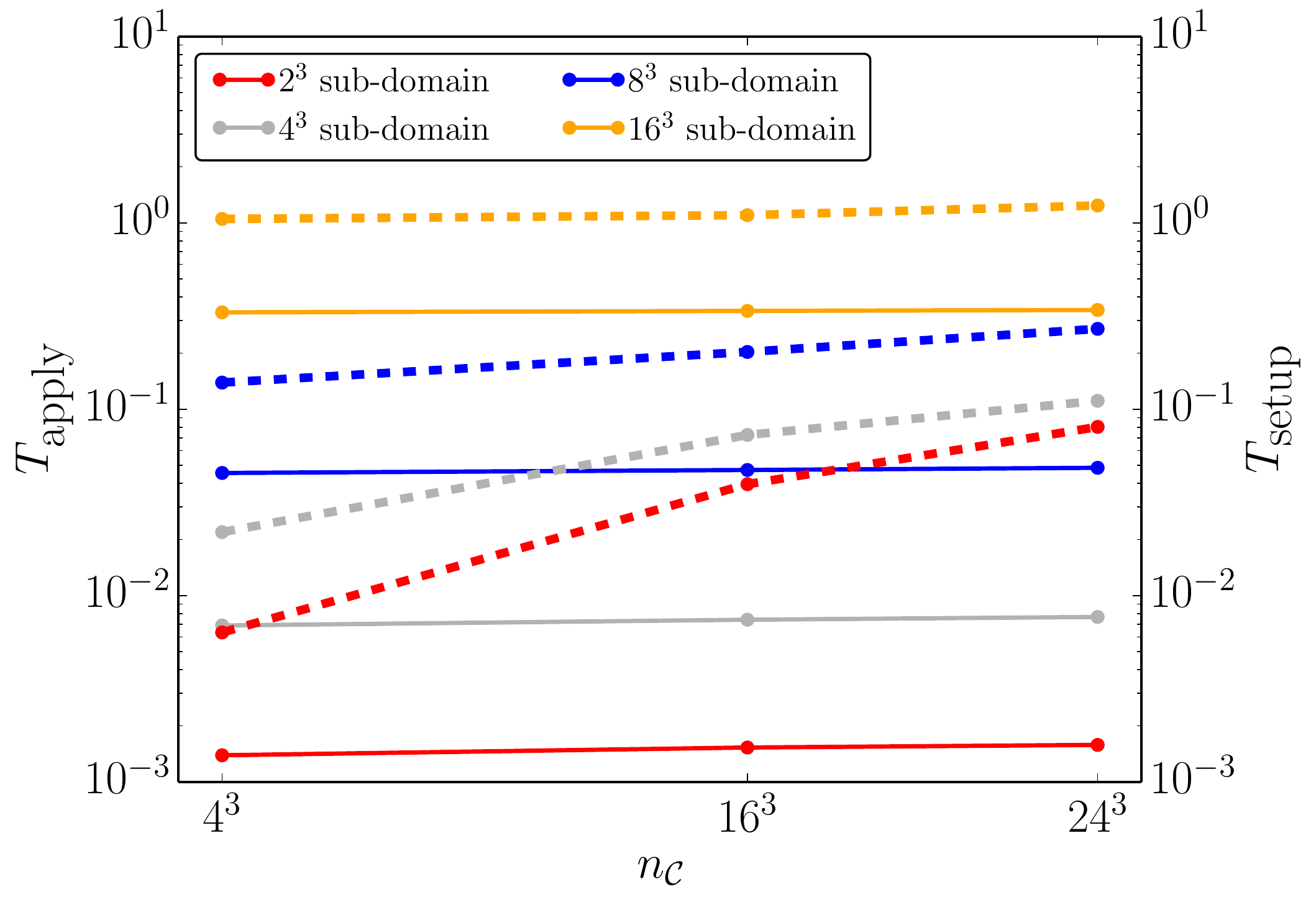}
\caption{Application time ($T_\mathrm{apply}$ -- left axis, solid lines) and setup time ($T_\mathrm{setup}$ -- right axis, dashed lines) as a function of 
the number of MPI-ranks ($n_{\mathcal C}$) for different sub-domain sizes using Disc. B.
A constant reduction factor of $r=8$ was used. \label{fig:agglomprof}}
\end{figure}

In Fig.~\ref{fig:agglomprof} we show the variation of the setup and application time with respect to the  number of elements 
within each sub-domain ($m^3$) using \emph{Disc. B} with different numbers of MPI-ranks ($n_\mathcal{C}$).
In these experiments, a fixed rank-reduction factor of $r = 8$ was used. 

From Tables~\ref{tab:telescopeA} and \ref{tab:telescopeB} we observe that for both stencils with low and high numbers of non-zero entries, 
the setup time required for repartitioning is observed to be only weakly dependent on $r$. 
In the worst case, reducing the size of the communicator by a factor of 64 is less than two times slower than if the communicator size was reduced by a factor of 8.
The time required for application of \pobj{Telescope} (vector permutation and inter-communicator scattering) is more strongly dependent on $r$. For instance, with a coarsening factor of 64, the application takes less than $5$ times longer compared to when $r=8$.
In all experiments, the setup time is always larger than the time required to apply the preconditioner.
Even when using approximately 32k cores, the setup time is less than 0.2 seconds.
From Fig.~\ref{fig:agglomprof} it is observed that application time of \pobj{Telescope} is independent of the size of the communicator and is only a function of the sub-domain size.
For a given sub-domain size, setup times appear to saturate, 
with small sub-domains ($2^3$) saturating at higher core counts compared to experiments using large sub-domains ($16^3$).

\subsection{Repartitioning at Scale} \label{sec:partitioningatscale}
To demonstrate the performance of \pobj{Telescope} in the context of a multigrid preconditioner, we 
consider the discrete solution of the 3D elasticity equations (Eq.~\eqref{eq:elasticity}) for displacement $\cvec u$ 
with $\cvec g = \cvec 0$ in a unit cube domain, $\Omega = [0,1]^3$. 
The constitutive behaviour of the elastic body is assumed to be isotropic, with uniform material properties 
$E= 1$ (Young's modulus) and $\nu = 0.33$ (Poisson ratio) throughout the domain.
Deformation is driven by the imposition of the following boundary conditions;
\begin{alignat*}{3}
\cvec u \cdot \cvec n &= 0,                              \qquad & \cvec t \cdot \cmat\sigma \cdot \cvec n &= 0,       \qquad & \text{for } z &= 0, \\
\cvec u \cdot \cvec n &= 2(\tfrac{1}{2} - x),       \qquad & \cvec t \cdot \cmat\sigma \cdot \cvec n &= 0,      \qquad & \text{for } x &= 0, 1, \\
\cvec u \cdot \cvec n &= 2(\tfrac{1}{2} - y),       \qquad & \cvec t \cdot \cmat\sigma \cdot \cvec n &= 0,      \qquad & \text{for } y &= 0, 1, \\
\cvec n \cdot \cmat \sigma \cdot \cvec n &= 0,  \qquad & \cvec t \cdot \cmat\sigma \cdot \cvec n &= 0,     \qquad & \text{for } z &= 1,
\end{alignat*}
where $\cvec n, \cvec t$ are the unit vectors normal and tangential to the boundary and $\cmat \sigma = \bar{\cmat C} \epsilon[\cvec u]$ is 
the Cauchy stress.

The domain $\Omega$ is discretized using $M^3$ $Q_2$ finite elements defined using the \pobj{DMDA} structured grid object 
which will be partitioned over the MPI communicator $\mathcal C$. 
Our finite-element implementation has the restriction that each sub-domain assigned to a given MPI-rank must contain at 
least one $Q_2$ element. This implementation restriction defines the depth of a geometric multigrid hierarchy 
which can be constructed on $\mathcal C$, 
and thus serves as our truncation strategy (see Sec.~\ref{sec:intro} and~\ref{sec:truncation}).

In these solver experiments, we use FGMRES preconditioned with a single V-cycle of geometric multigrid 
employing Galerkin coarse-level operators. Iterations are terminated when the initial residual is reduced by a factor of $10^{-8}$.
On the coarsest level, we use an inexact Krylov solve (GMRES preconditioned with block-Jacobi) which is terminated when the initial residual  
is reduced by a factor of $10^{-4}$. On all other levels, we used eight iterations of Richardson's method, preconditioned  
with Jacobi as the smoother. 

Our experiments consider the end-member scenario associated with strong scaling in which the 
number of elements per MPI-rank on the finest level is only $2^3$. The total number of elements in the mesh is 
given by $M^3$.
In Table~\ref{tab:telescopeMG} we report the setup time for \pobj{Telescope} ($T^{tele}_\text{setup}$) and 
solve time ($T_\text{solve}$) obtained using either a truncation strategy, 
or \pobj{Telescope} with different partitioning choices.
$T^{tele}_\text{setup}$ includes the time required for all phases of the setup, and multiple instances of \pobj{Telescope}.
When invoking \pobj{Telescope} in these experiments, the 
operator was required to be explicitly permuted during the setup phase and thus the time required for $\hat{\dmat P}^T \dmat A \hat{\dmat P}$ 
is included in the value reported for $T^{tele}_\text{setup}$. 
Given the size of the sub-domain considered, truncation variants of multigrid can utilize a hierarchy consisting of only two levels.
To obtain raw solve times without hidden setup costs, 
we report the walltime of the second of two consecutive (identical) solves.
All experiments were performed on Edison. 
\begin{table}[h!]
\centering
\caption{Multigrid performance. 
Shown are the setup time for $\mathtt{Telescope}$ ($T^{tele}_\mathrm{setup}$) and solve time ($T_\mathrm{solve}$) for
different meshes and multigrid configurations. 
The total number of different resolution meshes is indicated by $N_L$.
Hierarchical information related to repartitioning (``levels'' and ``ranks'') are separated by commas, 
ordered from the finest level (left) to the coarsest level (right).
}
\begin{tabular}{l l c l r r}
\toprule
$M$    &levels   &$N_L$  &ranks  &$T^{tele}_\text{setup}$ (s) &$T_\text{solve}$ (s)\\
\toprule
32 &2          &2 &$16^3$                       &--   &8.34E$-$01 \\ 
32 &2, 3      &4 &$16^3$, $4^3$           &8.56E$-$02   &5.23E$-$01 \\ 
32 &2, 3, 3  &6 &$16^3$, $4^3$, 1       &9.54E$-$02   &1.27E$-$01 \\ 
\midrule
64 &2          &2  &$32^3$                    &--                   &1.48E$+$01 \\ 
64 &2, 3      &4  &$32^3$, $8^3$            &2.30E$-$01                       &1.40E$-$01 \\ 
64 &2, 3, 3  &6  &$32^3$, $8^3$, $2^3$        &3.71E$-$01   &1.82E$-$01 \\ 
64 &2, 2, 3  &5  &$32^3$, $16^3$, $4^3$    &3.43E$-$01   &1.39E$-$01 \\ 
64 &2, 2, 3, 3 &7 &$32^3$, $16^3$, $4^3$, 1 &3.71E$-$01 &1.51E$-$01 \\ 
\bottomrule
\end{tabular}
\label{tab:telescopeMG}
\end{table}

Table~\ref{tab:telescopeMG} clearly highlights the importance of repartitioning for large-scale problems.
Even at 4096 cores, a single stage of repartitioning with $r=64$ yields a time-to-solution which is $1.6\times$ faster 
than the truncated  approach. At the scale of 32k cores, adopting two stages of repartitioning yields a time-to-solution 
which is approximately $106\times$ faster than the truncated approach. 
Note that the problem we examined here is ``easy'' in the sense that there are no spatial variations in the coefficients $E$ or $\nu$.
In situations where strong coefficient heterogeneities are present, we expect further improvements in the time-to-solution as the 
truncated coarse-level solve will become increasingly more difficult to converge.
Comparing the results from the two meshes which employed the deepest multigrid hierarchy (e.g. $M=32$: levels = $2,3,3$ cf. $M=64$: levels = $2,2,3,3$), we observe excellent weak scaling behaviour 
with respect to the solve time. 
Using 32k cores, introducing two stages of repartitioning causes the setup time to increase by at most a 
factor of $2.3$ above that required for the configuration using a single stage of repartitioning. 
The setup cost required by using two (or three) stages of repartitioning (last three rows of Table~\ref{tab:telescopeMG}) is larger than the time 
required for the entire solve on 32k cores. However, we note that the total time for each operation is less than half a second.
We also note that the time-to-solution does not always decrease as more stages of repartitioning are used.

\subsection{Hybrid CPU-GPU Sub-Domain Smoothers}
The trend in emerging and next-generation parallel computing architectures is the 
inclusion of co-processors (e.g.~GPUs or Xeon Phi) on each compute node. 
Whilst such technology brings a new level of on-node parallelism, it also complicates 
software development as (i) current discrete high-end GPUs do not share the same 
memory as the CPU (in the foreseeable future), 
and (ii) the form of parallelism is sufficiently different from the MPI model 
for which most large-scale simulation platforms have been designed to support. 
The development of optimal and scalable algorithms which map to such architectures is 
essential to enable the next generation of application software to fully exploit the floating point potential 
offered by hybrid co-processing compute nodes.


In this work we consider casting the multigrid smoother as a restricted additive Schwarz method (RASM) and mapping each sub-domain problem to the accelerator,
resulting in a multilevel preconditioner which can utilize many hybrid compute nodes. 
On each overlapping sub-domain, we perform traditional smoothing such as Chebyshev preconditioned by Jacobi, or Richardson's method  preconditioned by Jacobi.
The RASM preconditioner is used with a single Richardson iteration performed on the global problem. 
The Richardson iteration is performed on the original problem and thus serves as a synchronisation step between the 
sub-domain problems and the global problem. Experiments with such approaches can be found in~\cite{luo2011scalable}.
The  advantages of this preconditioner are that it avoids latencies in the following two ways: 
(i) The smoothing operation is local and does not require any inter-node messages to be sent via MPI; 
(ii) The application of RASM requires only one memory copy from the host to the device (and vice versa) for the input and output 
vectors, respectively.

Whilst the smoother may be faster to apply than a pure MPI+CPU implementation, it could possess worse 
smoothing characteristics as a result of the reduced number of synchronization points. 
However, in the limit of the overlap size equalling the number of smoothing iterations, both methods should be identical.
Clearly, a trade-off has to be made between the size of the overlap for each sub-domain, the number of local smoothing 
iterations to be performed, and the optimal coarsening factor between grid levels.

In Table~\ref{tab:MGtelescopeASM} we present experiments performed on Piz Daint 
to explore the potential of such hybrid preconditioners when 
applied to the finite-element discretization of the elasticity operator described in Sec.~\ref{eq:elasticity}. 
GPU support within PETSc is facilitated though the ViennaCL library \cite{Rupp:ViennaCL} with the OpenCL backend.
One design characteristic of the current PETSc-ViennaCL integration is that there is an assumed 
binding between a single MPI-rank and a GPU. However, to obtain the best per-node performance, we 
wish to use all cores together with the GPU. This is essential as only the sub-domain smoother is intended to be executed on the GPU. 
By using \pobj{Telescope} with a reduction factor of $r = 8$ and specifying RASM as the preconditioner for the sub-\pobj{KSP}, 
we can map the sub-domain problem to a single GPU via a single core.


\begin{table}[h!]
\centering
\caption{Setup and solve times for (i) a CPU only (upper four rows) and (ii) hybrid CPU-GPU RASM multigrid preconditioner. 
CPU-only experiments are denoted by an ``overlap''  of ``$-$''. Hybrid configurations which are faster than the CPU-only preconditioner 
are shown in bold.
}
\begin{tabular}{r r r r r r}
\toprule
$M$    &levels     &overlap     &$T_\text{setup}$ (s) &Its. &$T_\text{solve}$ (s)\\
\toprule
8	&2		&$-$			&1.12E$-$02		&12	&4.27E$-$02 \\
12	&3		&$-$			&4.41E$-$02		&16	&2.06E$-$01 \\
24	&3		&$-$			&1.88E$-$01		&13	&1.55E$+$00 \\
48	&4		&$-$			&1.29E$+$00		&11	&9.92E$+$00  \\
\midrule
\midrule
8	&2		&0			&5.49E$-$01		&12	&2.2813e-01 \\
12	&2		&0			&2.52E$+$00		&16	&2.3985e-01 \\
24	&3		&0			&4.94E$+$00		&13	&\textbf{1.28E$+$00} \\
48	&4		&0			&3.58E$+$01		&11	&\textbf{6.66E$+$00} \\
\midrule
8	&2		&1			&5.95E$-$01		&12	&2.40E$-$01  \\
12	&2		&1			&1.10E$+$00		&16	&4.30E$-$01  \\
24	&3		&1			&5.55E$+$00		&13	&\textbf{1.52E$+$00} \\
48	&4		&1			&2.30E$+$01		&11	&\textbf{7.34E$+$00} \\
\bottomrule
\end{tabular}
\label{tab:MGtelescopeASM}
\end{table}

All experiments used FGMRES preconditioned with a single V-cycle of geometric multigrid 
employing re-discretized coarse grid operators. 
Outer iterations are terminated when the initial residual is reduced by a factor of $10^{-8}$.
As a reference, we report computations with a standard CPU-only multigrid preconditioner and compare these results with 
the RASM - GPU sub-domain smoother (see upper four rows in Table~\ref{tab:MGtelescopeASM}).
The smoother used in both the CPU-only and the RASM hybrid method consists of Chebyshev(10) preconditioned with Jacobi.
The coarse-level solver consisted of one GMRES iteration preconditioned with block-Jacobi (defined over 64 MPI-ranks) and an
ILU(0) sub-domain solve. 
The RASM overlap is defined in terms of the number of $Q_2$ elements.
Assembled matrices were used on both the CPU and GPU.
All experiments were performed on Piz Daint using 8 nodes and all 8 cores per node (64 MPI-ranks in total).

From Table~\ref{tab:MGtelescopeASM} we observe that the iteration counts of the solve are not affected by using a local sub-domain smoother, 
hence we conclude that the hybrid RASM approach preserves the smoothing properties of the reference smoother (Chebyshev - Jacobi).
We also note that cases using an overlap larger than 1 also converged in the same number of iterations as their CPU-only counterpart, however 
all solve times were larger than using the CPU-only variant. The GPU-based RASM smoother is found to yield improved time-to-solution 
in several instances.

To understand this behaviour, in Table~\ref{tab:spmv} we report the per-node performance in terms of walltime and FLOP-rate (GF/s) and elements processed per second ($E/s$) achieved by the CPU (using 8 MPI-ranks) and GPU implementations of the sparse matrix vector product (SpMV). 
We observe that when the mesh contained more than $4^3$ elements, the walltime obtained using the GPU SpMV implementation is lower than that obtained using a CPU-only, flat-MPI SpMV.
The largest walltime ratio between the GPU and CPU is $\approx 2$. Ignoring all latencies associated with the Aries network, memory addressing and memory copies between the device and host, the best possible improvement in solve time which we can expect for this problem on Piz Daint is  thus $\approx 2\times$.
For the $M=48$ case with overlap zero, we obtained a solution $\approx 1.5\times$ faster than with the flat-MPI, CPU-only preconditioner.

We emphasize that it is expected that no speed-up was observed when using small sub-domains. For those cases, there is insufficient floating point work for the GPU to perform to offset the overhead associated with kernel launches and the memory copies passing through the PCI Express.

\begin{table}[h!]
\centering
\caption{Per-node performance of SpMV on Piz Daint.
Results were obtained from profiling 100 sparse matrix-vector products for the 
3D elasticity operator defined over $M^3$  $Q_2$ finite elements. 
}
\begin{tabular}{r r r r r r r}
\toprule
	   &\multicolumn{2}{c}{CPU (8 MPI-ranks)}	&\multicolumn{2}{c}{GPU} \\
$M$    &Time (s)     &GF/s   &$E/s$  &Time (s)     &GF/s &$E/s$\\
\toprule
4	&8.89E$-$03	&11.99	&720k &2.43E$-$02	&4.40 &264k\\
8	&1.27E$-$01	&6.96	&402k &5.90E$-$02	&14.99 &865k\\
12	&4.15E$-$01	&7.3		&417k &1.91E$-$01	&15.91 &908k\\ 
24	&3.15E$+$00	&7.79	&439k &1.44E$+$00	&17.09 &963k\\
\bottomrule
\end{tabular}
\label{tab:spmv}
\end{table}

\section{Discussion}
The agglomeration implementation used in \pobj{Telescope} adopts a top-down approach by 
providing support to coarsen an MPI communicator. 
In considering the development of general purpose agglomeration algorithms to use within solvers, 
this choice is most natural. The consequence is that \pobj{Telescope} is most useful in the context 
of mesh coarsening with a multilevel hierarchy. This is justified assuming the user defines a fine mesh 
with sub-domains which are well-balanced with respect to computational work and communication requirements. 
We acknowledge that our approach may lead to a small load imbalance on coarser levels. 
However, given the reduced volume of 
work to be performed on these levels, this can be tolerated. 

The philosophy used in \pobj{Telescope} differs from the typical approach adopted with unstructured meshes (e.g. UG4 \cite{reiter2013massively}), in which 
the multigrid hierarchy is created from mesh refinement, and thus the hierarchy of MPI 
communicators is constructed from \textit{refining} the size of the communicator assigned to the 
coarsest level (e.g.~MPI-ranks are added).
Such a strategy for constructing a hierarchy of MPI communicators is not general enough
to support the two use cases discussed in Sec.~\ref{sec:sscontsize} and~\ref{sec:ssdiffsd}. 
However, the latter scenario is highly specific to structured grids.

There are a number of memory optimizations which could be applied to the \pobj{Telescope} implementation. 
Specifically, during the setup phase of the preconditioner, at least one temporary matrix has to be held in memory on the communicator $\mathcal C$.
In the event that the matrix is to be explicitly permuted, an additional temporary matrix must also be stored.
Future work should focus on removing the number of temporary matrices required during the setup phase.
Nevertheless, given that the typical use case of \pobj{Telescope} is the redistribution of matrices in which there are very few unknowns per MPI-rank, this storage overhead is not critical.

Currently \pobj{Telescope} only supports repartitioning \pobj{DMDA}s. 
An obvious extension is to provide support for a wider range of \pobj{DM} implementations, 
particularly \pobj{DMPlex}, which is the most general mesh representation object provided by PETSc. 
Extending support to \pobj{DMPlex} likely mandates introducing a new \pobj{DM} interface to permit ``refining'' the MPI communicator. Such an interface would be of the form \pobj{DMRefinePartition(DM dm,MPI\_Comm rcomm,IS *perm,DM *rdm)}, where \pobj{dm} is the original mesh, \pobj{rcomm} is the refined MPI communicator, \pobj{rdm} is the repartitioned \pobj{DM} object and \pobj{perm} is an index-set defining the permutation between the original and repartitioned DOF ordering.

Lastly, in the context of the multilevel preconditioners discussed in this work, 
there is currently no performance model to guide the optimal selection of when agglomeration 
should occur, and or what the rank reduction factor should be. 
Future work should automate these choices based on machine characteristics (e.g. network latency, memory access costs), together 
with characteristics of the matrix (e.g. number of non-zeros per row). An agglomeration performance model is required to support this development.

\section{Summary}
We have presented an implementation of process agglomeration (or MPI communicator coarsening) 
called \pobj{Telescope} which has been introduced into the PETSc library. Whilst the development of \pobj{Telescope} was 
motivated by the need to have a scalable coarse-level solver in the context of a multigrid preconditioner, the design of 
this agglomeration component is sufficiently general to allow it to be used in many other contexts.  
Through a series of numerical experiments related to (i) the end-member of a strong scaling study and (ii) a hybrid 
smoother which utilizes both CPUs and GPUs, we have demonstrated the benefits of this agglomeration 
implementation. 

\section{Acknowledgments}
The Swiss National Supercomputing Centre (CSCS) and NERSC are thanked for compute time on Piz Daint and Edison respectively.
PS acknowledges financial support from the Swiss University Conference and the Swiss Council of Federal
Institutes of Technology through the Platform for Advanced Scientific Computing (PASC) program.
Some of the authors were supported by the U.S. Department of Energy, Office of Science, Advanced Scientific Computing Research under Contract DE-AC02-06CH11357.

{\bf Government License.} The submitted manuscript has been created by UChicago 
Argonne, LLC,
Operator of Argonne National Laboratory (``Argonne'').
Argonne, a U.S. Department of Energy Office of Science laboratory, is
operated under Contract No. DE-AC02-06CH11357. The U.S. Government
retains for itself, and others acting on its behalf, a paid-up
nonexclusive, irrevocable worldwide license in said article to reproduce,
prepare derivative works, distribute copies to the public, and perform
publicly and display publicly, by or on behalf of the Government.


\begin{thebibliography}{10}

\bibitem{Adams:2004:UIF:1048933.1049978}
M.~F. Adams, H.~H. Bayraktar, T.~M. Keaveny, and P.~Papadopoulos.
\newblock Ultrascalable implicit finite element analyses in solid mechanics
  with over a half a billion degrees of freedom.
\newblock In {\em Proceedings of the 2004 ACM/IEEE Conference on
  Supercomputing}, SC '04, pages 34--, Washington, DC, USA, 2004. IEEE Computer
  Society.

\bibitem{petsc-user-ref}
S.~Balay, S.~Abhyankar, M.~F. Adams, J.~Brown, P.~Brune, K.~Buschelman,
  L.~Dalcin, V.~Eijkhout, W.~D. Gropp, D.~Kaushik, M.~G. Knepley, L.~C.
  McInnes, K.~Rupp, B.~F. Smith, S.~Zampini, and H.~Zhang.
\newblock {PETS}c users manual.
\newblock Technical Report ANL-95/11 - Revision 3.6, Argonne National
  Laboratory, 2015.

\bibitem{petsc-web-page}
S.~Balay, S.~Abhyankar, M.~F. Adams, J.~Brown, P.~Brune, K.~Buschelman,
  L.~Dalcin, V.~Eijkhout, W.~D. Gropp, D.~Kaushik, M.~G. Knepley, L.~C.
  McInnes, K.~Rupp, B.~F. Smith, S.~Zampini, and H.~Zhang.
\newblock {PETS}c {W}eb page.
\newblock \url{http://www.mcs.anl.gov/petsc}, 2015.

\bibitem{petsc-efficient}
S.~Balay, W.~D. Gropp, L.~C. McInnes, and B.~F. Smith.
\newblock Efficient management of parallelism in object oriented numerical
  software libraries.
\newblock In E.~Arge, A.~M. Bruaset, and H.~P. Langtangen, editors, {\em Modern
  Software Tools in Scientific Computing}, pages 163--202. Birkh{\"{a}}user
  Press, 1997.

\bibitem{blatt2012massively}
M.~Blatt, O.~Ippisch, and P.~Bastian.
\newblock A massively parallel algebraic multigrid preconditioner based on
  aggregation for elliptic problems with heterogeneous coefficients.
\newblock {\em arXiv preprint arXiv:1209.0960v2}, 2012.

\bibitem{brown2013achieving}
J.~Brown, B.~Smith, and A.~Ahmadia.
\newblock Achieving textbook multigrid efficiency for hydrostatic ice sheet
  flow.
\newblock {\em SIAM Journal on Scientific Computing}, 35(2):B359--B375, 2013.

\bibitem{emans2011coarse}
M.~Emans.
\newblock Coarse-grid treatment in parallel {AMG} for coupled systems in {CFD}
  applications.
\newblock {\em Journal of Computational Science}, 2(4):365--376, 2011.

\bibitem{gmeiner2014parallel}
B.~Gmeiner, H.~K{\"o}stler, M.~St{\"u}rmer, and U.~R{\"u}de.
\newblock Parallel multigrid on hierarchical hybrid grids: a performance study
  on current high performance computing clusters.
\newblock {\em Concurrency and Computation: Practice and Experience},
  26(1):217--240, 2014.

\bibitem{gmeiner2015performance}
B.~Gmeiner, U.~R{\"u}de, H.~Stengel, C.~Waluga, and B.~Wohlmuth.
\newblock Performance and scalability of hierarchical hybrid multigrid solvers
  for {S}tokes systems.
\newblock {\em SIAM Journal on Scientific Computing}, 37(2):C143--C168, 2015.

\bibitem{gmeiner2015towards}
B.~Gmeiner, U.~R{\"u}de, H.~Stengel, C.~Waluga, and B.~Wohlmuth.
\newblock Towards textbook efficiency for parallel multigrid.
\newblock {\em Numerical Mathematics: Theory, Methods and Applications},
  8(01):22--46, 2015.

\bibitem{hoefler2013mpi}
T.~Hoefler, J.~Dinan, D.~Buntinas, P.~Balaji, B.~Barrett, R.~Brightwell,
  W.~Gropp, V.~Kale, and R.~Thakur.
\newblock {MPI + MPI}: a new hybrid approach to parallel programming with {MPI}
  plus shared memory.
\newblock {\em Computing}, 95(12):1121--1136, 2013.

\bibitem{isaac.sisc.2015}
T.~Isaac, G.~Stadler, and O.~Ghattas.
\newblock Solution of nonlinear {S}tokes equations discretized by high-order
  finite elements on nonconforming and anisotropic meshes, with application to
  ice sheet dynamics.
\newblock {\em SIAM Journal on Scientific Computing}, 37(6):B804--B833, 2015.

\bibitem{lechmann2011comparing}
S.~M. Lechmann, D.~A. May, B.~J.~P. Kaus, and S.~M. Schmalholz.
\newblock Comparing thin-sheet models with {3-D} multilayer models for
  continental collision.
\newblock {\em Geophysical Journal International}, 187(1):10--33, 2011.

\bibitem{li2007use}
J.~Li and O.~B. Widlund.
\newblock On the use of inexact subdomain solvers for {BDDC} algorithms.
\newblock {\em Computer Methods in Applied Mechanics and Engineering},
  196(8):1415--1428, 2007.

\bibitem{luo2011scalable}
L.~Luo, C.~Yang, Y.~Zhao, and X.-C. Cai.
\newblock A scalable hybrid algorithm based on domain decomposition and
  algebraic multigrid for solving partial differential equations on a cluster
  of {CPU/GPU}s.
\newblock In {\em 2nd International Workshop on GPUs and Scientific
  Applications (GPUScA 2011)}, page~45, 2011.

\bibitem{may2015scalable}
D.~A. May, J.~Brown, and L.~Le~Pourhiet.
\newblock A scalable, matrix-free multigrid preconditioner for finite element
  discretizations of heterogeneous {S}tokes flow.
\newblock {\em Computer Methods in Applied Mechanics and Engineering},
  290:496--523, 2015.

\bibitem{mcinnes2014hierarchical}
L.~C. McInnes, B.~Smith, H.~Zhang, and R.~T. Mills.
\newblock Hierarchical {K}rylov and nested {K}rylov methods for extreme-scale
  computing.
\newblock {\em Parallel Computing}, 40(1):17--31, 2014.

\bibitem{muller2014massively}
E.~H. M{\"u}ller and R.~Scheichl.
\newblock Massively parallel solvers for elliptic partial differential
  equations in numerical weather and climate prediction.
\newblock {\em Quarterly Journal of the Royal Meteorological Society},
  140(685):2608--2624, 2014.

\bibitem{reiter2013massively}
S.~Reiter, A.~Vogel, I.~Heppner, M.~Rupp, and G.~Wittum.
\newblock A massively parallel geometric multigrid solver on hierarchically
  distributed grids.
\newblock {\em Computing and Visualization in Science}, 16(4):151--164, 2013.

\bibitem{RokosGorman13}
G.~Rokos and G.~Gorman.
\newblock {PRAgMaTIc}--parallel anisotropic adaptive mesh toolkit.
\newblock In R.~Keller, D.~Kramer, and J.-P. Weiss, editors, {\em Facing the
  Multicore-Challenge III}, volume 7686 of {\em Lecture Notes in Computer
  Science}, pages 143--144. Springer Berlin Heidelberg, 2013.

\bibitem{rudi2015extreme}
J.~Rudi, A.~C.~I. Malossi, T.~Isaac, G.~Stadler, M.~Gurnis, P.~W. Staar,
  Y.~Ineichen, C.~Bekas, A.~Curioni, and O.~Ghattas.
\newblock An extreme-scale implicit solver for complex {PDEs}: highly
  heterogeneous flow in {E}arth's mantle.
\newblock In {\em Proceedings of the International Conference for High
  Performance Computing, Networking, Storage and Analysis}, page~5. ACM, 2015.

\bibitem{Rupp:ViennaCL}
K.~Rupp, F.~Rudolf, and J.~Weinbub.
\newblock {ViennaCL - {A} high level linear algebra library for {GPUs} and
  multi-core {CPUs}}.
\newblock In {\em International Workshop on GPUs and Scientific Applications},
  pages 51--56, 2010.

\bibitem{triangle:homepage}
J.~{Shewchuk}.
\newblock {Triangle: {A} two-dimensional quality mesh generator and Delaunay
  triangulator}.
\newblock \url{http://www-2.cs.cmu.edu/~quake/triangle.html}, 2005.

\bibitem{shewchuk96}
J.~R. Shewchuk.
\newblock Triangle: {E}ngineering a {2D} quality mesh generator and {D}elaunay
  triangulator.
\newblock In M.~C. Lin and D.~Manocha, editors, {\em Applied Computational
  Geometry: Towards Geometric Engineering}, volume 1148 of {\em Lecture Notes
  in Computer Science}, pages 203--222. Springer-Verlag, May 1996.
\newblock From the 1$^\text{st}$ ACM Workshop on Applied Computational
  Geometry.

\bibitem{tetgen:homepage}
H.~{Si}.
\newblock {TetGen: {A} quality tetrahedral mesh generator and three-dimensional
  Delaunay triangulator}.
\newblock \url{http://tetgen.berlios.de}, 2005.

\bibitem{Si2015}
H.~Si.
\newblock {TetGen}, a {Delaunay}-based quality tetrahedral mesh generator.
\newblock {\em ACM Transactions on Mathematical Software (TOMS)}, 41(2),
  February 2015.

\bibitem{sundar2012parallel}
H.~Sundar, G.~Biros, C.~Burstedde, J.~Rudi, O.~Ghattas, and G.~Stadler.
\newblock Parallel geometric-algebraic multigrid on unstructured forests of
  octrees.
\newblock In {\em Proceedings of the International Conference on High
  Performance Computing, Networking, Storage and Analysis}, page~43. IEEE
  Computer Society Press, 2012.

\bibitem{trottenberg2000multigrid}
U.~Trottenberg, C.~W. Oosterlee, and A.~Schuller.
\newblock {\em Multigrid}.
\newblock Academic Press, 2000.

\end{thebibliography}
\end{document}